\documentstyle[preprint,aps]{revtex}
\begin{document}
\preprint{BNL-CAP-\phantom{12345}}
\draft
\title{Muon Transverse Ionization Cooling:\protect\\
Stochastic Approach}
\author{ R. C. Fernow and J. C. Gallardo}
\address{ Center for Accelerator Physics\protect\\
Brookhaven National Laboratory\protect\\
Upton, New York 11973}
\date{\today}
\maketitle
\begin{abstract}
Transverse ionization cooling of muons is modeled as a Brownian motion  of the
muon beam as it traverses a Li or Be rod. A Langevin like equation is written
for the  free particle case (no external transverse magnetic field) and
for the case of a harmonically bound beam in the presence of a focusing
magnetic field. We demonstrate that the well known muon cooling
equations for short-absorbers can be extrapolated to the useful case of a
long absorber rod with a focusing magnetic field present.
\end{abstract}
\section{INTRODUCTION}
The possibility of a $\mu^+\mu^-$ collider to explore the Higgs energy range
and supersymmetry has begun to be vigorously examined. One of the crucial
issues to achieve the required luminosity (${\cal L}\approx 10^{34}\,cm^{-2}
s^{-1}$) is the
need to compress the phase space by means of muon cooling. A technique that has
been shown to be very promising is ionization cooling. The introduction of the
concept and the physics was first discussed by Skrinsky\cite{ref0}; for a clear
and comprehensive treatment we refer the reader to Neuffer's
articles\cite{ref1}.

The original derivation in ref.\cite{ref1} of the transverse cooling
differential equation assumed a cooling system consisting
of small alternating absorber and reaccelerator sections. Subsequently, Palmer
\cite{ref2} and Fernow\cite{ref3} have argued that the equation is valid for a
single long absorber. Their argument is quite straightforward; from the
definition
of emittance it is possible to show that
\begin{eqnarray}
{d\epsilon_{\perp}^N\over dz}&=&- |{dE_{\mu}\over dz}|{\epsilon_{\perp}^N
\over \beta^2E_{\mu}} +
\frac{1}{2}{(\gamma \beta )\over \epsilon_{\perp}^N(z)}\left[
<r^2>{d<\theta^2>\over dz}+<\theta^2>{d<r^2>\over dz}
\right]\nonumber \\
& - &{(\gamma\beta)\over \epsilon_{\perp}^N(z)}<r\theta>{d<r\theta>\over dz}
\label{eq1}
\end{eqnarray}
where we have used ${d(\gamma\beta)\over dz} ={1\over\beta mc^2}
{dE_{\mu}\over dz},$
 $\epsilon_{\perp}^N$ is the muons' normalized transverse emittance, $E_{\mu}$
is the muon total energy, $\beta_{\perp}$ is the beta function and
$<r^2>,$ $<\theta^2>$ are the square of the rms position and divergence of the
beam due to multiple scattering.
 The first term in Eq.\ref{eq1} reflects the energy loss ({\it cooling})
and the last
three terms are produced by multiple scattering ({\it heating}).

Assuming a long cooling rod (Li or Be) the gaussian approximation is quite
adequate, then
\begin{eqnarray}
<y^2>&=&\frac{1}{3}\theta_c^2 z^3 \nonumber\\
<\theta^2> &=&\theta_c^2z
\label{eq3}
\end{eqnarray}
where the projected scattering angle
$\theta_o\equiv \theta_c \sqrt{z}= {13.6 [MeV] \over \beta cp}
\sqrt{{z\over L_R}}$ and $L_R$ is the radiation length.  In this expression
we have neglected
logarithmic correction terms\cite{ref4}. There is  also, the more accurate
formula  for $\theta_o$\cite{ref5}
\begin{equation}
\theta_o={\chi_c\over \sqrt{1+F^2}} \sqrt{{1+\nu\over \nu}\log{(1+\nu)}-1}
\end{equation}
with the characteristic angle $\chi_c= {\sqrt{0.157 [MeV]z}\over \beta
p}\sqrt{{Z(Z+1)\over A}};$ the phenomenological parameters
$F=0.98$ and $\nu= {\Omega_o \over 2(1-F)}$ are chosen to fit the
experimental data ($\Omega_o$ represent the mean number of scatters in the
medium).

 If we neglect from Eq.\ref{eq1} the third and fourth terms, then we can write
 the minimum achievable emittance as\cite{ref1},
\begin{equation}
\epsilon_{\perp}^N|_{min}={(13.6[MeV])^2 \over 2 \beta m c^2 |{dE_{\mu}\over
dz}|}
{\beta_{\perp}(0)\over L_R} \label{eq5}
\end{equation}
 Palmer\cite{ref2} has pointed out that the third and fourth terms in
Eq.\ref{eq1} need not be included because the multiple scattering medium (Li or
Be rod) is immersed in a uniform transverse magnetic field, which prevents the
beam
from  spreading laterally. This raises the question, how is the functional form
of the particle distribution changed in position and angle  due to the external
magnetic field?
We will examine this question in the next sections.

We should also point out that the treatment considered here is also of interest
for a number of other problems in accelerator physics. These include scattering
of particles by residual gas in a synchrotron\cite{ref6}, scattering in the
Inverse \u{C}erenkov
accelerators\cite{ref9}, plasma beat-wave accelerators\cite{ref10} and plasma
lenses for future linear colliders\cite{ref11} and the dynamics of space-charge
dominated beams\cite{ref12}.

\section{PARTICLE DISTRIBUTION WITHOUT A MAGNETIC FIELD}

This problem has been analyzed in detail by Rossi\cite{ref7} following the
Fokker-Planck equation approach. Let ${\cal W}(y,\theta,z;y_o,\theta_o)\, dy
\,d\theta$ represent the number of particles in the phase space element
$(y, y+dy;
\theta, \theta+d\theta)$ after traversing a medium of thickness z and initial
coordinates $y(0)=y_0$, $\theta(0)=\theta_o.$ It satisfies
the equation
\begin{equation}
{\partial {\cal W}\over \partial z}=-(\theta -\Theta_o){\partial {\cal W}\over
\partial y}+
{\theta_c^2\over 2} {\partial^2 {\cal W}\over \partial \theta^2} \label{eq6}
\end{equation}
with boundary conditions ${\cal
W}(y,\theta,z;y_o,\theta_o)|_{z=0}=\delta(y-y_o)\delta(\theta-\theta_o)$
and  solution
\begin{equation}
{\cal W}(y,\theta,z:y_o,\theta_o)={\sqrt{3} \over \pi z^2
\theta_c^2}\exp{\left [ -{2\over \theta_c^2}\left ({(\theta-\Theta_o)^2\over z}
-{3(y-Y_o)(\theta-\Theta_o)\over z^2}+{3(y-Y_o)^2\over z^3}\right )\right
]}\label{eq7}
\end{equation}
which can be verified by direct substitution with $\Theta_o=\theta_o$ and
$Y_o=y_o+\theta_oz.$
This result allows us to compute the emittance of the beam after traversing the
cooling rod. After tedious gaussian integrations we obtain
\begin{eqnarray}
<y>&=& Y_o\nonumber \\
<\theta >&=& \theta_o\nonumber \\
<y^2> &=&Y_o^2+ {\theta_c^2 z^3\over 3}\nonumber\\
<\theta^2>&=&\Theta_o^2+\theta_c^2 z\nonumber\\
<y\theta>&=&\Theta_o Y_o+{\theta_c^2 z^2\over 2}\label{eq8}
\end{eqnarray}
Averaging over the initial coordinates and assuming gaussian distributions with
standard deviations $\sigma_{yo}$ and $\sigma_{\theta o},$ we
obtain
\begin{eqnarray}
\ll y\gg&=&\ll \theta \gg=0\nonumber\\
\ll y^2\gg &=&\sigma_{yo}^2+\sigma_{\theta o}^2z^2+2z< y_o\theta_o >+
{\theta_c^2 z^3\over 3}\nonumber\\
\ll\theta^2\gg&=&\sigma_{\theta o}^2+\theta_c^2 z\nonumber\\
\ll y\theta\gg&=&\sigma_{\theta o}^2 z+< y_o\theta_o>+{\theta_c^2 z^2\over 2}
\label{eq9}
\end{eqnarray}
The total emittance in the absence of a focusing field is
\begin{equation}
\epsilon_{\perp}(z)=\sqrt{\epsilon_{\perp}^2(0) +{\theta_c^4z^4\over
12} +\sigma_{\theta o}^2\theta_c^2 {z^3\over
3}+\theta_c^2< y_o\theta_o> z^2+\sigma_{y o}^2\theta_c^2z }\label{eq10}
\end{equation}
The terms proportional to $\theta_c$ are the contributions due to multiple
scattering.

\section{PARTICLE DISTRIBUTION WITH A MAGNETIC FIELD}

Consider now the problem of a particle traversing a rod of material that has an
axial current flowing through it; such a particle satisfies the equation of
motion ${d^2y\over dz^2} +K(z)y =0$ where $K(z)={e B \over mc \gamma\beta
a}=\omega^2,$
$B$ is the azimuthal magnetic field and $a$ is the radius of the rod. $K(z)$
is a function of z because of the energy loss, but
 for the simplicity of the arguments that follow, we neglect the energy change
as the
beam traverses the rod.

A more complete treatment must take into account random accelerations of the
particles due to
scattering (i.e. stochastic changes in angle ${dy\over
dz}=\theta$). A correct equation of motion is
\begin{equation}
{dy\over dz}=\theta \quad , \quad {d\theta\over dz}+K(z)y =A(z) \label{eq12}
\end{equation}
where we denote with $A(z)$ the random acceleration due to Coulomb scattering
which excites betatron oscillations in the beam.
This equation is formally a Langevin equation of a particle in an external
field K(z)y (harmonic oscillator) where the frequency is a function of the {\it
time} variable $z.$ The main assumptions regarding the stochastic variable
$A(z)$, more precisely $\int_z^{z+dz} dz' A(z'),$ is that it is independent
of $y,$ that it varies
extremely rapidly compared to the variations of the coordinates $y$ and
$\theta,$ and that it is Gaussian-distributed with a variance $\theta_c^2.$

Therefore, casting the muon cooling problem in stochastic terms, we first
determine the
particle distribution ${\cal W}(y,\theta,z;y_o\theta_o)$, and then as before,
 we  calculate the emittance from
that function.

The method for determining the distribution function uses standard techniques
for  solving ordinary differential equations and is described in detail by
Chandrasekhar\cite{ref8}. The result for the distribution function is,
\begin{eqnarray}
\lefteqn{{\cal W}(y,\theta,z,\omega;y_o,\omega_o) =  }\nonumber \\
&&{1\over 2\pi \sqrt{\{
FG-H^2\}}}\exp{\left[-{1\over 2 \{ FG-H^2\}} \left( G(y-Y_o)^2
-2H(y-Y_o)(\theta -\Theta_o)
+F(\theta -\Theta_o)^2\right) \right]}
\label{eq14}
\end{eqnarray}
where the parameters F,G H are functions of the external focusing field;
\begin{eqnarray}
F&=&\theta_c^2 {z\over 2\omega^2}(1- {\sin{2\omega z}\over 2\omega
z})\nonumber\\
G&=&\theta_c^2 {z\over 2}(1+ {\sin{2\omega z}\over 2\omega z})\nonumber\\
H&=&\theta_c^2 {1\over 2\omega^2}\frac{1}{2}(1- \cos{2\omega z})\nonumber\\
FG-H^2&=& {\theta_c^4\over \omega^4}z^2 \left[ 1-
({\sin{\omega z}\over \omega z})^2\right]\label{eq15}
\end{eqnarray}
It can be shown that Eq.\ref{eq14} reproduces Eq.\ref{eq7} in the limit $\omega
\rightarrow 0,$ and the probability density ${\cal W}(y,\theta,z,\omega;
y_o,\theta_o)$ satisfies a
parabolic partial differential equation, the Fokker-Planck equation
\begin{equation}
{\partial{\cal W}\over \partial z}=-(\theta-\Theta_0){\partial {\cal W}\over
\partial y}+
\omega^2 (y-Y_o){\partial {\cal W} \over \partial \theta}+\frac{1}{2}
\theta_c^2
{\partial^2{\cal W}\over \partial \theta^2}\label{eq16}
\end{equation}
 As in the previous section we are interested in calculating the
second moments of the distribution and from those the emittance. We find that
\begin{eqnarray}
<y>&=&y_o \cos{\omega z} +\theta_o z {\sin{\omega z}\over \omega z}\nonumber \\
<\theta>&=&\theta_o \cos{\omega z} -y_o\omega \sin{\omega z}\nonumber\\
<y^2>&=&(y_o \cos{\omega z} +\theta_o z {\sin{\omega z}\over \omega z})^2+
      z {\theta_c^2 \over 2 \omega ^2} (1- {\sin{2\omega z}\over 2\omega z})
\nonumber\\
<\theta^2>&=&(\theta_o \cos{\omega z} -y_o\omega \sin{\omega z})^2+
      z {\theta_c^2 \over 2} (1+ {\sin{2\omega z}\over 2\omega z})\nonumber\\
<y\theta >&=&y_o\theta_o \cos{2\omega z}-{y_o^2\omega \over 2}\sin{2\omega
z}+\theta_o^2 z\cos{\omega z}{\sin{\omega z}\over \omega z}+{\theta_c^2 z^2
\over 2}({\sin{\omega z}\over \omega z})^2\label{eq17}
\end{eqnarray}
If we now assume an uncorrelated ensemble of incident particles with
independent gaussian distributions
of initial conditions $y_o$ and $\theta_o$ and  average over both variables, we
obtain
\begin{eqnarray}
\ll y\gg&=&\ll \theta \gg=0\nonumber\\
\ll y^2\gg&=&\sigma_{yo}^2 (\cos{\omega z})^2 +\sigma_{\theta o}^2
z^2 ({\sin{\omega z}\over \omega z})^2+
      z {\theta_c^2 \over 2 \omega ^2} (1- {\sin{2\omega z}\over 2\omega z})
\nonumber\\
\ll\theta^2\gg&=&\sigma_{\theta o}^2 (\cos{\omega z})^2 +\sigma_{y o}^2
\omega^2
(\sin{\omega z})^2+
      z {\theta_c^2 \over 2} (1+ {\sin{2\omega z}\over 2\omega z})\nonumber\\
\ll y\theta \gg&=&\sigma_{\theta o}^2 z
{\sin{2\omega z}\over 2\omega z}-\sigma_{yo}^2 \frac{1}{2}\omega \sin{2\omega
z}+{\theta_c^2 z^2
\over 2}({\sin{\omega z}\over \omega z})^2\label{eq18}
\end{eqnarray}

An important observation is that for a very high magnetic field ($\omega>>0$)
the rms beam size remains approximately constant. If the beam
is focused to a waist at the entrance to the rod, the emittance at any distance
z inside the rod is\cite{ref15}
\begin{equation}
\epsilon_{\perp}^2(z) = \epsilon_{\perp}^2(0)+\sigma_{yo}^2\theta_c^2 z+
{\theta_c^4\over 4 \omega^2} z^2+ {\theta_c^4\over 8 \omega^4} \left[
1-\cos{(2\omega z)}\right]\label{eq20}
 \end{equation}
The second term gives the dominant contribution of multiple scattering to the
emittance.
This also leads to Eq.\ref{eq5} for the minimum emittance, provided that
$ \sigma_{yo}^2\gg {\theta_c^2\over 2 \omega^2} L_{rod}$ and
$ \sigma_{yo}^2\gg {\theta_c^2\over 4 \omega^3},$ which will usually be the
case for strong focusing ($\omega \gg 0$).

\section {CONCLUSIONS}
Using an analogy with a random dynamical process modelled with a Langevin
equation, we have incorporated the stochastic nature of both
position and angle variables into the problem of a muon traversing a Li or
Be rod immersed in
a uniform azimuthal magnetic field. The pseudo-Brownian motion of the
particles in the medium represents heating of the beam. The emittance increase
due
to  multiple Coulomb scattering (the less likely single and plural scattering
events are neglected) opposes  the emittance decrease (cooling),
introduced by the energy loss ${dE_{\mu}\over dz}.$ With sufficiently strong
focusing present, ionization cooling can effectively take place over long
lengths of
absorbing material.

\section*{ACKNOWLEDGEMENTS}
We wish to thank D. Neuffer for a critical reading of the manuscript.
This research was supported by the U.S. Department of Energy
under Contract No. DE-ACO2-76-CH00016.
\newpage

\newpage
\end{document}